\documentstyle[mprocl]{article}

\begin{document}

\author{STEVEN WEINSTEIN}

\address{Department of Philosophy, Northwestern University\\
1818 Hinman Ave, Evanston, IL  60208-1315, USA}

\title{TIME, GAUGE, AND THE SUPERPOSITION PRINCIPLE IN QUANTUM GRAVITY}

\maketitle\abstracts{
The quantization of time-reparametrization invariant systems such as
general relativity is plagued by an ambiguity relating to the role of time
in the theory. If one parametrizes observables by the (unobservable) time,
and then relies on the existence of an approximate ``clock'' degree of
freedom to give physical meaning to the observables, one finds multiple
quantum states that yield the same predictions yet interfere with each other.%
}

General relativity admits a kind of Hamiltonian formulation,
in which the Hamiltonian generates change in the geometry of a spacelike
hypersurface with respect to an arbitrary ``time''
parameter $\tau $.
The arbitrariness of the time-evolution is reflected in the
fact that the Hamiltonian is the sum of the supermomentum and
super-Hamiltonian {\em constraints}. This presents no difficulty in
principle for the classical theory, but gives rise to various ``problems of
time'' in the quantum theory. (See Isham~\cite{Ish93} and Kucha\v{r}~\cite{Kuc92a}.)

In this talk, we will examine the quantization of the parametrized
non-relativistic particle. This system is much simpler than general
relativity, and indeed it is frequently invoked as a simple example of the 
{\em efficacy }of the Dirac constraint-quantization method for
time-reparametrization invariant systems. However, we will find that
not only does the Dirac method fail to return us in any
straightforward way to ordinary Schr\"{o}dinger quantum mechanics, but it gives
rise to a theory in which there are distinct quantum states with apparently
identical physical content, the superposition of which yields a state with
different physical content.

\section{Classical parametrized particle}

Consider a non-relativistic particle moving in three dimensions. If we
treat $t$ as a dynamical variable, and parametrize the motion by an arbitrary
parameter $\tau=\tau(t)$, we find that we have a
constraint 
\begin{equation}
{\cal H}=N(p_{t}+H)=0\,\mbox{ ,}  \label{constraint}
\end{equation}
where $N=dt/d\tau $, $p_{t}$ is the ``momentum'' canonically conjugate to $%
t $, and $H=H(x_{i},p_{j})\,$is the usual Hamiltonian.%
~\footnote{See Unruh \& Wald~\cite{UW89} for a more thorough development
of the parametrized formalism.}
This constraint is
called the ``super-Hamiltonian'' and it generates change with respect to the
parameter $\tau $ via the equations of motion 
\[
\frac{dx_{i}}{d\tau }=\{x_{i},{\cal H\}}=N\frac{\partial H}{\partial p_{i}}\mbox{ , }
\frac{dp_{i}}{d\tau }=-\{p_{i},{\cal H}\}=-N\frac{\partial H}{\partial
x_{i}} 
\]
and 
\begin{equation}
\frac{dt}{d\tau }=\{t,{\cal H}\}=N\mbox{ , }\frac{dp_{t}}{d\tau }=-\{p_{t},
{\cal H}\}=0\,\mbox{ .}
\end{equation}

\section{Constraint quantization}

The ``constraint quantization'' method~\cite{Dir65} involves turning the
classical constraints into operators, and imposing them as constraints on
the allowed state-vectors. Thus the first step in the constraint
quantization of this system is to solve the equation $\widehat{{\cal H}}\Psi
(x^{i},t)=0$. If we define $\hat{t}:=t$ and $\hat{p}_{t}:=-i\hbar \frac{\partial }{\partial t}$,
this equation takes the form 
\begin{equation}
i\hbar \frac{\partial }{\partial t}\Psi (x^{i},t)=\widehat{H}\Psi (x^{i},t)%
\mbox{ ,}
\end{equation}
which looks rather like the Schr\"{o}dinger equation. It is not---the wave
functions are functions of both $x^{i}\,$and $t$. Note, too, that the
equation {\em only} takes the functional form of the Schr\"{o}dinger equation if
one represents $\hat{t}$ by the multiplication operator. These issues aside,
solving the equation is straightforward---one finds that solutions are of
the form $\Psi (x^{i},t)=e^{-(i/\hbar )\ \widehat{H}t}\circ \psi (x^{i})$.

The next step is to turn these solutions into a Hilbert space. A useful way
of determining the inner-product is the algebraic method of Ashtekar and
Tate~\cite{AT94};
one selects a complete algebra of ``observables'' on the
solution space, and one requires that the inner-product be such that these
observables are self-adjoint. One choice is 
\[
\widehat{X}^{i}(0)\circ \Psi :=\widehat{U}(0)\hat{x}^{i}\widehat{U}%
^{-1}(0)\circ \Psi =e^{-(i/\hbar )\widehat{H}t}\hat{x}^{i}\circ \psi \equiv
e^{-(i/\hbar )\widehat{H}t}\circ x^{i}\psi (x^{i}) 
\]
and 
\begin{equation}
\widehat{P}_{i}(0)\circ \Psi :=\widehat{U}(0)\hat{p}_{i}\widehat{U}%
^{-1}(0)\circ \Psi =e^{-(i/\hbar )\widehat{H}t}\hat{p}_{i}\circ \psi \equiv
e^{-(i/\hbar )\widehat{H}t}\circ -i\hbar \frac{\partial }{\partial x^{i}}%
\psi (x^{i})\,\mbox{ ,}
\end{equation}
where $\widehat{U}(0):=e^{-(i/\hbar )\widehat{H}t}\,$. These observables
intuitively correspond to the position and momentum at some time $t=0$.
Requiring them to be self-adjoint gives an inner-product of 
\begin{equation}
\left\langle \Psi (x^{i},t),\Phi (x^{i},t)\right\rangle =\int \Psi
^{*}(x^{i},t),\Phi (x^{i},t)\ dx\mbox{ .}
\end{equation}

\section{Observables}

We see that one can straightforwardly construct one-parameter families of observables
$\widehat{X}^{i}(t)$ and$\,\widehat{P}_{i}(t)$ that correspond, intuitively,
to position and momentum at different times. This allows one to talk about
the value of an observable at any given time $t$, as in ordinary quantum
theory. However, in ordinary quantum theory, the time $t$ is an external
parameter corresponding to the classical time in which the system is
embedded. If we are to consider the parametrized particle as an analogue for
general relativity, then we cannot think about time in this way---there is
no ``external environment'' in which the system is embedded. What, then, is
the physical content of the theory?\ Given a state $\Psi (x^{i},t)$, one can
determine the expectation value for $X^{1}$ at some time $t$, but this is
useless if one does not have a way of physically ascertaining the time,
which is, after all, not an observable.

Even in ordinary quantum theory, one doesn't measure time directly.
One uses a clock, and time is determined, e.g., by the position of the
hands of the clock. In this spirit, let us assume there are good
``clock'' variables available that give one physical, observable access to
the time, so that if one knows the state $\Psi (x^{i},t)$, and one knows the
time, one can find expectation values for the remaining observables.

Suppose that $\widehat{X}^{3}(t)$ is such a ``clock'' variable.\footnote{
I.e., $X^{3}$ is classically a monotonic function of $t$, and
its ``quantum fluctuations'' are small enough that it is highly
improbable that it will be observed to ``run backward.'' 
Although such ``good enough''
clocks arguably suffice for coarse-grained predictions, the known absence of any quantum analogue of a perfect clock~\cite{UW89}
may render this scheme useless at the Planck scale, which is of course the very scale at which quantum-gravitational
effects are expected to come into play.}
If it is a
good clock variable, it will establish fairly reliable correlations with the
Schr\"{o}dinger time $t$, and allow us to give some operational meaning to
observables $\widehat{X}^{i}(t)$ and $\widehat{P}_{i}(t)\;$(where $i$ now
runs from $1$ to $2$) parametrized by $t$. But now consider a wave-function $%
\Phi (x^{i},t)=e^{-5(i/\hbar )\widehat{H}}\Psi (x^{i},t)$, corresponding
to a simple displacement of $\Psi $ by five units of time. This
wave-function yields exactly the {\em same correlations} between the clock
variable $\widehat{X}^{3}(t)$ and all of the other observables. One would
therefore like to say that it represents the {\em same} physical {\em system}%
, just as one would be inclined to say (along with Leibniz) that translating
the entire classical world five feet to the left in Newtonian absolute space
yields the same universe.
The choice between $\Phi $ and $\Psi $ looks like a choice of ``gauge'', which is not
so surprising given that the Hamiltonian takes the form of a constraint.
However, the two states 
interfere, and so their superposition $\alpha
\Phi +\beta \Psi $ does {\em not} yield the same correlations. Thus we have
a situation in which two physically equivalent states may be superimposed to
yield a different state.

\section{Conclusion}

The superposition principle appears to fail for the constraint-quantized
parametrized particle, and it would appear that constraint-quantized general
relativity is subject to the same problem. Whereas in ordinary quantum
theory, the superposition of physically equivalent states (i.e., states
differing by a phase) yields a physically equivalent state, this is not the
case for parametrized systems. The breakdown stems from the lack of a
fiducial external observer or reference system.
Lacking an external reference to give independent physical meaning to $t$,
one must fall back on internal correlations, and this
leads to the failure of the superposition principle.

\end{document}